\documentclass[%
 reprint,
nofootinbib,
 amsmath,amssymb,
 aps, prd,
floatfix,
showkeys,
]{revtex4-2}

\usepackage{graphicx}
\usepackage{dcolumn}
\usepackage{bm}
\usepackage{amsmath}
\usepackage{slashed}

\begin{document}

\preprint{}

\title{Inflation and Dark Matter from The Low Entropy Hypothesis and Modeling Mechanism of Modified Gravity}

\author{Jackie C.H. Liu}
 \email{chjliu@connect.ust.hk, jackieliu@astri.org, jackieliu0@gmail.com}%

\affiliation{%
Hong Kong Applied Science and Technology Research Institute, Hong Kong
}%
\affiliation{%
Department of Physics, Hong Kong University of Science and Technology, Hong Kong
}%

\date{\today}

\begin{abstract}

The hypothesis of low entropy in the initial state of the universe usually explains the observed entropy increase is in only one time direction: the thermodynamic arrow of time. 
The Hamiltonian formalism is commonly used in the context of general relativity.  The set of Lagrange multipliers are introduced in the formalism, and they are corresponding to the Hamiltonian constraints which are written in terms of "weak equality" - the equality is satisfied if the constraints hold. 
Follow the low-entropy hypothesis, we postulate a modeling mechanism - a \textit{weak equality} (of modeling) that holds only on the subspace of the theory space of physical models defined by some \textit{modeling constraints}.
By applying the modeling mechanism, we obtain a specific model of modified gravity under specific modeling conditions.
We derive a novel equation of modeling from the mechanism, that describes how different gravitational models emerge.
The solution of the modeling equation naturally turns out to be the model of $R^2$-gravity (with additional terms) if ordinary matter is negligible.
We also found that this mechanism leads to two models: large-field inflation and wave-like dark matter.
Interestingly, the wave-like dark matter model is supported by the most recent observations of Einstein rings.

\end{abstract}

\maketitle

\section{\label{sec:level1}INTRODUCTION}

Big Bang theory has been widely accepted since the discovery of the cosmic microwave background (CMB) \cite{Smooth2007}.  However, there are puzzles regarding the homogeneity from the observations of CMB and the fine-tuning of the initial condition: the horizon problem and flatness problem.
The hypothesis of cosmic inflation is a commonly accepted solution \cite{Wang2014}.  It not only solves the problems of horizon and flatness, but also implies primordial fluctuation spectra by quantum fluctuations.  Such fluctuations are indeed observed with high precision \cite{Smooth2007,planck2018}.

Observations from both early universe and local measurements agree that dark matter is a major component of matter content under general relativity (GR) formalism \cite{1310.6061}.   Local measurements such as galaxy rotation curves support the existence of dark matter (DM).  However, there is an apparent disagreement between the dark-matter-particle picture, such as core-cusp halos, and an overabundance of satellite dwarf galaxies.  Different DM models, such as wave-like dark matter, try to avoid the cusp problem and explain the flatness of the rotation curve.

Bose-Einstein condensed scalar field dark matter (SFDM) is an active model to be studied \cite{1310.6061}.  The system described by Klein-Gordon and Einstein field equations seems to fit the observational constraints of the cosmic microwave background (CMB) and the abundances of the light elements produced by big-bang nucleosynthesis (BBN).  There are a number of studies on the wave nature of DM, such as ultralight bosons, a soliton core comprising a coherent standing wave at the center of halos, and self-interfering waves \cite{1510.07633,1902.10488, 2101.11735, 0801.1442}.  Most recently, in Ref. \cite{2304.09895}, the authors even suggested a wave-like dark-matter model can reproduce these aspects, whereas particle DM fails by analyzing the observation of gravitational lensing.

There is a camp of modified gravities (MG) instead of the hypothesis of the existence of DM to explain the \textit{apparent anomaly} from GR, such as MOND, TeVeS, $f(R)$, general higher-order theories \cite{1106.2476}.
In the Sakharovian approach, the Einstein-Hilbert action is simply the first approximation of a much more complicated action/theory.  In addition, Kellogg showed that these theories are renormalizable in the presence of matter fields at the one-loop level \cite{stelle1977}. 

In this study, we propose a mechanism for the modeling of (classical) field theories, which leads to both large-field inflation and wave-like dark matter from the underlying modified gravity model after conformal transformation.  The motivation for the construction of the mechanism is based on the hypothesis of low-entropy condition in the universe's initial state. 
According to Goldstein et al. \cite{PhysRevD.94.023520}, "to account for the observed entropy, increase in only one time direction, one usually appeals to the hypothesis that the initial state of the universe was one of very low entropy."  
(Goldstein et al. also analyzed different cosmological models, which lead to the emergence of a thermodynamic arrow of time without assuming the hypothesis of low entropy.)  In our work, by assumption of the low-entropy hypothesis, we propose there is a mechanism that respects low-entropy hypothesis, and it leads to models' generation. 

The constraints of the Hamiltonian formalism of GR are written as "weak equality", which holds only on the subspace of the phase space \cite{PhysRevD.103.024032}.
In Section 2, 
we introduce the \textit{modeling mechanism}  using weak equality (of modeling), and show how the generated modeling constraints can be applied to the modeling of gravity theories.
We speculate that there is a specific state of universe described by the weak equality of modeling, such that a specific model of gravity is generated.

In Section 3, under the $f(R,\phi)$ formalism developed in Refs. \cite{Harko_2013}, we obtain modeling equations describing how the MG varies, such that different gravitational models might emerge.
We found that this solution led to two interesting gravitational models.

In Section 4, from the first model of the solution, we obtain the well-known $R^2$-gravity (with additional terms) from the solution of the modeling equation if the ordinary matter is negligible.\footnote{
Starobinsky found that theories with $R^2$ correction might lead to an early period of de Sitter expansion. The spectrum of scalar and tensor fluctuation generated during this type of inflation have been well studied, and they are compatible with observations of the CMB \cite{1106.2476}.
}  Then, we apply the conformal transformation, and the generated model is equivalent to general relativity with a scalar field in the Einstein frame.  We found that the slow roll condition was satisfied by such a model, with the e-fold being approximately $\approx 61$.  This is consistent with previous observations \cite{planck2018}.  We also calculate the slow roll parameters, $\epsilon_V, \eta_V$, at the crossing, which are consistent with observations as well.
We found that the modeling mechanism operates in the Jordan frame and generates modified gravity, while the effect of modified gravity disappears in the Einstein frame, with a new degree of freedom as the scalar field is obtained.  

In Section 5, we derive the dark matter model from the second model of the solution.  In the Einstein frame, the generated scalar field is effectively equivalent to a Klein-Gordon scalar field, which leads to a wave-like dark-matter scenario.  
From Refs. \cite{1310.6061,0801.1442, 2304.09895}, wave-like dark matter ($\psi$DM) might offer an explanation for the dark matter paradigm to fit large-scale measurements (SFDM behaving as collision-less on dust) such as CMB, large-scale structure, and local-scale measurements (avoiding the cusp problem owing to the wave nature of DM).

Therefore, we postulate that the mechanism might naturally lead to the model of inflation and wave-like dark matter.

Finally, from the solution of the mechanism, we find a polynomial (relating the power of the logarithm of the Ricci scalar) from the constraints.  In Section 6, we briefly study the possibility of the energy-scale hierarchy implied in the Jordan frame by the solution of the mechanism.

\section{Low Entropy Principle and Modeling Mechanism}

The hypothesis about a low entropy initial state of the Universe is studied in \cite{PhysRevD.94.023520}.  We call it low-entropy hypothesis.
The observed entropy increase is in only one time direction; therefore, low-entropy hypothesis is naturally assumed such as the study.
In \cite{carroll2004spontaneous,CARROLL_2005}.
Carroll and Chen's models, along with the work by Barbour et al., offer a potential explanation for the universe's low entropy state in the remote past. Instead of merely assuming, like the past hypothesis does, that the initial state of the universe was exceptionally unique, these models provide an alternative explanation without relying solely on that assumption. 
There exists a possibility that a fundamental principle exists, which naturally leads to a state of low entropy. Building upon the study of the low-entropy hypothesis, we put forward a principle that aligns with this hypothesis and subsequently construct a modeling mechanism in this section.

The cosmic microwave background observations provide evidence that the early universe reached a state of thermal equilibrium \cite{Smooth2007}, where the principles of Boltzmann entropy were applicable,
\begin{equation}
	\mathcal{S} = k_B \log{\Omega}, 
\end{equation}
where $\mathcal{S}, k_B, \Omega$ are entropy, Boltzmann's constant, and the number of possible micro states of a system respectively.
We propose \textit{the low entropy principle} (LEP) as 
\begin{quote}
	\textit{The initial state of the universe is in a specific condition that minimizes Boltzmann's entropy.}
\end{quote}
The introduction of Boltzmann's entropy in LEP to describe the special initial state of the universe implicitly assumes that each particle has an identical independent probability distribution, and interactions/correlations between the particles are negligible.
If there is a special condition of the initial state of the universe which is effectively described by a single field, this condition trends to minimize the possible micro states ($\Omega$) and (Boltzmann's) entropy; in such special condition, interactions among fields are effectively irrelevant.  
We propose there is a mechanism which respects LEP and leads to effectively single-field-dependent system at a specific condition. We found the mechanism generating models simultaneously,
\footnote{In this work, we study the generating of models of modified gravity under the context of $F(R,\phi)$ formalism}
 so we name such mechanism the \textit{modeling mechanism}.
Therefore, we conjecture that the generation of models is the consequence of the modeling mechanism in the initial state of the universe.

In this rest of this section, we first introduce the weak equality of modeling.  Then we construct the modeling mechanism described by the weak equality.

Hamiltonian constraints are commonly used in the context of GR \cite{PhysRevD.103.024032}.  The Hamiltonian with Arnowitt-Deser-Misner (ADM) formalism is constructed from the lapse function, shift vector, induced metric, and their respective conjugate momenta.  The formalism introduces the set of Lagrange multipliers corresponding to the constraints written in "weak equality", such as $\mathcal{H}_i \approx 0_i, \pi_i \approx 0_i$, where $\pi_i$ denotes conjugate momenta, $\mathcal{H}_i \approx 0_i$ is the momentum constraint, and subscript $i$ denotes the spatially index.
"$\approx$" represents a weak equality that holds only in the subspace of the phase space defined by the constraints.

Following the notation of the weak equality in Hamiltonian constraints, we postulate a \textit{weak equality} (of modeling) that holds only on the subspace of the theory space of physical models defined by some \textit{modeling constraints}.
In this section, we show how to obtain the weak equality (of modeling) and modeling constraints from the variation of action.
The basic idea of the modeling mechanism is to propose a modeling function that can be used to extract the modeling constraints and connected to the variation of action.

Assume that there are multiple real scalar fields ($\phi_i$) for the models in the theory space, and one of scalar fields is called as the \textit{base field}.
The \textit{modeling function} $\mathcal{E}$ is defined as follows:
\begin{equation}
	\mathcal{E} := \sum^n_{k=0} F_k \; O_k[\phi_0] \approx 0, \label{eq:ModelFn}
\end{equation}
where 
$\phi_0$ and $F_k$ denote the base field and form factors (some functions), respectively, and $O_k$ denotes the combinations of the base field (and its derivatives, such as $\phi_0,\phi_0 ^ 2,\phi_0 ^ 3, \partial_x \phi_0$...)
Weak equality means that the modeling function is equal to zero when some modeling constraints (to be identified) hold, and such modeling constraints define the subspace of models (in model space), that is, the particular theories.
The motivation for the modeling function in Eq. (\ref{eq:ModelFn}), is that given the arbitrariness of the base field, the weak equality holds only if all form factors are zero, that is, $F_k = 0$.
The modeling constraints are simply a set of equations, $F_k = 0$, if  the models are sufficiently determined by those modeling constraints.

To follow the low entropy principle, a single-field-dependent system at a specific condition is obtained such that the possible micro state ($\Omega$) of the initial state of the universe is minimized. We propose the modeling mechanism described by a weak equality as
\begin{equation}
   \delta S \approx \delta S[\phi_0],    \label{eq:WE}
\end{equation}
where $\delta S$  and $\delta S[\phi_0]$ denote the variation of the action depending on all fields and the base field, respectively, and $\mathcal{E}$ is naturally obtained.
In this study, we propose that there is a \textit{specific state} of the universe in which
	\textit{theories/models are generated by the weak equality in Eq. (\ref{eq:WE})}.  Furthermore, we deduce the physical conditions/constraints in such a state.
We show how weak equality (\ref{eq:WE}) generates the modeling constraints.  Let us expand the weak equality (\ref{eq:WE}) to the sum of the terms of the individual variation of fields as follows:
\begin{equation}
   \sum^m_{i=0} \int d^4x (\hat E_i \mathcal L) \delta \phi_i \approx  \int d^4x' (\hat E_0 \mathcal L) \delta \phi_0,
\end{equation}
where $\hat E_i$ denotes the \textit{Euler-Lagrange operators} obtained when applying the variation with respect to fields $\phi_i$ (for instance, $\hat E_0 = \frac{\partial}{\partial \phi_0} - \frac{d}{dt} \frac{\partial}{\partial \dot\phi_0} $ for the action, $S[\phi_0(t)]$); and $d^4x' $ denotes the differential elements of the integral at the R.H.S., which is not necessarily equal to that in the L.H.S., that is, $d^4x' \ne d^4x $.  By definition, the weak equality is valid when the modeling constraints hold.  If we impose the modeling constraints (of the base field), $\delta\phi_i = h_i(x) \delta\phi_0$, for some smooth functions $h_i(x)$, we have
\small
\begin{equation}
   \sum^m_{i \neq 0} \int d^4x (\hat E_i \mathcal L) h_i(x) \delta\phi_0+ \int d^4x (\hat E_0 \mathcal L) \delta \phi_0
   \approx  \int d^4x' (\hat E_0 \mathcal L) \delta \phi_0.
\end{equation}
\normalsize  
By allowing the proportionality relationship of the differential elements, $d^4x' = c \; d^4x $, where there is a non-zero constant $c$, we obtain,
\footnote{Note the Euler-Lagrange operator is invariant for the transformation of $ dt' \to c\; dt$.}
\begin{equation}
    \int d^4x \sum^m_{i \neq 0} (\hat E_i \mathcal L) h_i(x) \delta\phi_0
   \approx  \int d^4x (c-1) (\hat E_0 \mathcal L) \delta \phi_0.
\end{equation}
Therefore, by refering to Eq. (\ref{eq:ModelFn}), we infer that the modeling function $ \mathcal{E}$ can be obtained from the variation of the action, and should satisfy
\begin{equation}
   \mathcal{E} := \sum^n_{k=0} F_k \; O_k[\phi_0] = \sum^m_{i \neq 0} (\hat E_i \mathcal L) h_i(x)  +  (1-c) (\hat E_0 \mathcal L)
,  \label{eq:defE}
\end{equation}
such that the weak equality in Eq. (\ref{eq:WE}) holds true when $ \mathcal{E} \approx 0$.  
The R.H.S. of the equation above can be expanded according to different combinations of $\phi_0$ and derivatives of $\phi_0$ from the terms $(\hat E_i \mathcal L)$ so the form factors $F_k$ and expressions $O_k$ in the L.H.S. can be found accordingly.
In other words, if we require 
a modeling function that satisfies Eq. (\ref{eq:defE}) for the arbitrariness of the base field $\phi_0$, the modeling constraints ($F_k=0$), obtained.  Therefore, the modeling mechanism is constructed by introducing the weak equality of modeling (\ref{eq:WE}), modeling function (\ref{eq:defE}), and modeling constraints (of the base field), $\delta\phi_i = h_i(x) \delta\phi_0$. 
We found that the modeling mechanism generated specific models of modified gravity (Section 3).

According to the definition of the modeling function in Eq. (\ref{eq:ModelFn}), the number of modeling constraints (to be generated) must be sufficient to determine the models.
We consider conditions that lead to a sufficiently enough number of constraints.

Let us write the Lagrangian in the form by only ordering through combinations of the base field, as follows: 
\begin{equation}
\mathcal L = \sum^N_{r=0} p_r(\phi_1,\phi_2,...; \partial_\mu \phi_1, \partial_\mu \phi_2, ...) O_r[\phi_0],
  \label{eq:LagModel1}
\end{equation}
where $p_r$ denotes the model parameters/functions, and the number of DOF  of the model (degrees of freedom of the model parameters/functions) is explicitly equal to $N+1$. 
From Eq. (\ref{eq:defE}), we obtain:
\begin{align*}
   \sum^n_{k=0} F_k \; O_k[\phi_0] = \sum^m_{i = 0} (\hat E_i \mathcal L) h_i(x) = \\
    \sum^N_{r=0}( \sum^m_{i = 0}  h_i(x) \hat E_i  (p_r O_r[\phi_0]))
\end{align*}
where we define $h_0(x) := (1 - c)$.  
To extract the form factors $F_k$, we apply the Euler-Lagrange operator $\hat E_i $ on terms $p_r O_r$ and substitute each non-base field in terms of the base field by $\phi_i = \phi_i(\phi_0)$.
\footnote{If we refer to the previously modeling constraints (of base field), $\delta\phi_i = h_i(x) \delta\phi_0$, which implies the dependencies of the non-base fields to base field, so we may assume such dependencies are smooth.  One can express such dependencies as, $\phi_i = \phi_i(\phi_0)$.}
The number of terms per each $i$'th term may increase according to the product rule of differentiation and/or higher order of differentiation because of differential operators, such as $\frac{d}{dt} \frac{\partial}{\partial \dot\phi_0}$.
After grouping the terms according to each $O_k[\phi_0]$ from the lower equation, we can extract the form factors, $F_k$.
Because of the Euler-Lagrange operators, the total sum of terms for each $O_k[\phi_0]$ is generally more than $N+1$; therefore, the total number of modeling constraints ($F_k=0$) is generally larger than or equal to the number of DOF, $N+1$.
We conclude that the model is sufficiently determined (or over-determined), because the form in Eq. (\ref{eq:LagModel1}).

Next, we postulate that the modeling mechanism applies to the context of the MG, that is, models of the MG generated by the modeling function obtained by WE (\ref{eq:WE}).
We refer to \cite{Harko_2013} for the work of $f(R,\phi)$ formalism of the MG.
Let us assume that the action depends on both the base field and the metric field; thus, WE (\ref{eq:WE}) reads
\begin{equation}
   \delta S[\phi_0, g_{\mu\nu}] \approx \delta S[\phi_0].
\end{equation}
Following the same argument for obtaining the modeling function, if we impose specific constraints (of the base field) corresponding to the metric field, $\delta g^{\mu\nu} = h^{\mu\nu}(x) \delta \phi_0$, for some smooth functions $h^{\mu\nu}(x)$, the weak equality becomes
\begin{align}
   \int d^{4}x \; \sqrt{-g} ( \hat E_{\mu\nu} \mathcal{L}) (h^{\mu\nu}(x) \delta \phi_0)+ \int d^4x \; \sqrt{-g} (\hat E_0 \mathcal L) \delta \phi_0 \nonumber \\
   \approx  \int d^4x' \sqrt{-g} (\hat E_0 \mathcal L) \delta \phi_0,
\end{align}
where $\hat E_{\mu\nu}$ denotes the Euler-Lagrange operators for the inverse metric field $g^{\mu\nu}$ and $g=det(g_{\mu\nu})$.
Note that the term $\hat E_{\mu\nu} \mathcal{L}$ above yields the gravitational field equation by  $\hat E_{\mu\nu} \mathcal{L}=0$.
Therefore, following Eq. (\ref{eq:defE}), we infer that the modeling function should satisfy:
\begin{equation}
   \mathcal{E} = \sum^3_{\mu=0} \sum^3_{\nu=0} ( \hat E_{\mu\nu} \mathcal{L}) h^{\mu\nu}(x)  +  (1-c) (\hat E_0 \mathcal L).   \label{eq:modelFn2}
\end{equation}
The modeling function described above is similar to that in Eq. (\ref{eq:defE}), except that the non-base fields are components of the metric field.

From the study of sufficient conditions, the number of terms of the modeling function per combination of base fields leads to a number of form factors/modeling constraints.  The sum of the indices ($\mu, \nu$) in Eq. (\ref{eq:modelFn2}) may lead to too many modeling constraints such that the model is over-determined.  To avoid this ambiguity, we follow the form of the model in Eq. (\ref{eq:LagModel1}), we assume that the form of Lagrangian as follows: 
\begin{equation}
	\mathcal L = \sum^N_{r=0} p_r(R) O_r[\phi_0],    \label{eq:LagModel2}
\end{equation}
where $R$ denotes Ricci scalar.  We found if
\begin{equation}
	h^{\mu\nu}(x) = c_0 \; g^{\mu\nu},  \label{eq:hCondition}
\end{equation}
where $c_0$ is a non-zero real constant
\footnote{$c_0$ is allowed to be non-real constant, however, the related study is beyond the scope of this work.}
, and we have a simple modeling function as 
\begin{equation}
   \mathcal{E} = c_0 (g^{\mu\nu} \hat E_{\mu\nu} \mathcal{L})  +  (1-c) (\hat E_0 \mathcal L), \label{eq:simpleModelFn}
\end{equation}
where $g^{\mu\nu} \hat E_{\mu\nu} \mathcal{L} = 0$ is nothing but the contracted gravitational field equation.  Therefore, the number of terms for obtaining form factors from the modeling functions is reduced owing to the terms summed by the contracted gravitational field equation and the simplified form of the Lagrangian, Eq. (\ref{eq:LagModel2}).
Finally, considering the identity, $\delta R = R_{\mu \nu} \delta g^{\mu \nu}$, Eq. (\ref{eq:hCondition}) and imposed constraints, $\delta g^{\mu\nu} = h^{\mu\nu}(x) \delta \phi_0$, we found the specific modeling constraint (of the base field to Ricci scalar) as,
\begin{equation}
	 R = R_0 \; e^{c_0( \phi_0 - v)},    \label{eq:fieldMapR}
\end{equation}
where we introduce the constants $R_0$ and $v$ for the integration constant.
As mentioned earlier in this section, we seek physical conditions/constraints in the specific state of the modeling mechanism described by WE (\ref{eq:WE}).  Eq. (\ref{eq:fieldMapR}) is the specific physical constraint obtained.  We call Eq. (\ref{eq:fieldMapR}), the \textit{field-map constraint}, because it correlates the mapping between the fields of $\phi_0$ and $R$.

In a later section, we solved the model parameters/functions $p_r (R)$, in the Lagrangian described in Eq. (\ref{eq:LagModel2}), and we show the modeling mechanism leading to specific MG models, that is, the MG theory generated in such a specific state of the modeling mechanism.
Furthermore, we found that if we substitute the base field $\phi_0$ with the Ricci scalar $R$ using the field-map constraint, Eq. (\ref{eq:fieldMapR}) into Lagrangian in Eq. (\ref{eq:LagModel2}), we obtain a specific model of gravity (in the context of $f(R)$ gravity), which is discussed in Section 4.
The scalar-field dependence of the model in Eq. (\ref{eq:LagModel2}), is coupled to the Ricci scalar because of the substitution of the field-map constraint in Eq. (\ref{eq:fieldMapR}).
We propose a model of gravity leading to the inflaton generated as a consequence of such a state.
In summary, the postulate is that
\textit{there is a specific state of universe (described by WE (\ref{eq:WE}))-the specific model of gravity generated while the scalar field ($\phi_0$) is 'absorbed' simultaneously.}
In this study, we consider this possibility.  Using conformal transformation, we postulate that the inflaton and dark matter models are generated in the Einstein frame according to the two solutions suggested by the mechanism in Sections 4 and 5.

\section{The Modified Gravity from Modeling Mechanism}

In this section, we apply the modeling mechanism introduced in the previous section to model the theories of gravity in $f(R,\phi)$ formalism by \cite{Harko_2013}.
First, we postulate a generic form of the Lagrangian and apply $f(R,\phi)$ formalism to obtain field equations of the metric and scalar fields.  From the modeling function, Eq. (\ref{eq:simpleModelFn}), the modeling constraints are derived by matching terms by terms of all orders of $\phi$ and all derivative terms of $\phi$.  We found two modeling equations from the constraints: one for theories of gravity and one for theories of the scalar field.  Finally, we solved the modeling equations to obtain the model of gravity and reproduce the well-known $R^2$ gravity.

Using the field equations from $f(R,\phi)$ formalism of Harko et al. \cite{Harko_2013}, we introduce a generic form
\footnote{We make use of the canonical kinetic term of scalar field $\phi$.}
 of the total Lagrangian with matter-minimally coupled, $\mathcal{L}_T$,
\begin{eqnarray}
	\mathcal{L}_T = \mathcal{L}_M +  \mathcal{L}_\text{gra} + \mathcal{L}_\phi, 
\end{eqnarray}
where
\begin{eqnarray}
	\mathcal{L}_\text{gra} = \frac{R}{2} g(R), \nonumber \\
	\mathcal{L}_\phi = \frac{1}{2} (\nabla \phi)^2 - V(R,\phi),  \nonumber 
\end{eqnarray}
and the subscripts $M$, $\text{gra}$, and $\phi$ denote the matter, gravitational, and $\phi$ sectors, respectively, and $(\nabla \phi)^2 = g^{\mu \nu} \nabla_\mu \phi \nabla_\nu \phi$.  Note when $g(R)=1$ and the absence of scalar field ($\phi$) implies that the gravitational model is simply general relativity.  We obtained the following field equations:\footnote{We use the natural unit of reduced Planck mass, $M_{pl}=1$ in this work, unless otherwise stated.} 
\begin{widetext}
\begin{eqnarray}(\mathcal{L}_T)_R
-\frac{1}{2} \left(\mathcal{L}_{\text{gra}}+\mathcal{L}_{\phi }\right) g_{\mu \nu }+ g_{\mu \nu } \nabla _{\sigma}\nabla^\sigma (\mathcal{L}_T)_R+(\mathcal{L}_T)_R R_{\mu \nu }-\nabla _{\mu }\nabla _{\nu }(\mathcal{L}_T)_R -\frac{1}{2} \tau _{\mu \nu }+\frac{1}{2}\nabla _{\mu }\phi \nabla _{\nu }\phi =0  \label{eq:MGFE0},\\
E_g=-2 \left(\mathcal{L}_{\text{gra}}+\mathcal{L}_{\phi }\right)+3\nabla _{\mu }\nabla^\mu (\mathcal{L}_T)_R+(\mathcal{L}_T)_R R -\frac{1}{2} \tau +\frac{1}{2} \nabla^\mu \phi \nabla _{\mu }\phi=0, \label{eq:MGFE1}\\
E_1=\frac{1}{\sqrt{-\mathit{g}}}\partial _{\mu }\left((\mathcal{L}_T)_{(\nabla \phi )^2}\sqrt{-\mathit{g}}g^{\mu \nu }\partial _{\nu }\phi\right)-\frac{(\mathcal{L}_T)_\phi}{2}=0,  \label{eq:MGFE2}
\end{eqnarray}  
\end{widetext}
where $\tau _{\mu \nu }$ is the energy momentum tensor of matter, $\tau=\tau_\mu^\mu$, $\mathit{g}$ is the determinant of metric $g_{\mu \nu }$; $(\mathcal{L}_T)_R, (\mathcal{L}_T)_\phi$, $(\mathcal{L}_T)_{(\nabla \phi )^2}$, $E_g=0$, and $E_1=0$ denote $\partial\mathcal{L}_T/\partial R, \partial\mathcal{L}_T/\partial \phi, \partial\mathcal{L}_T/\partial ((\nabla \phi )^2)$, the contract field equation of gravity, and the equation of motion of scalar field $\phi$ respectively.
The second equation is the contraction of the first equation above.  The field equations can be found in $f(R,\phi)$ formalism \cite{Harko_2013}.

By definition, the modeling function in Eq. (\ref{eq:simpleModelFn}) satisfies the weak equality of modeling, $\mathcal{E} \approx 0$, which leads to the correspondence of both equations of motion for the metric and scalar fields discussed, $E_g \propto E_1$, such that
\begin{eqnarray}
-\frac{(\mathcal{L}_T)_{\phi }}{2}-\frac{\sqrt{-g}\; \partial _{\mu }\left(\sqrt{-g} \;(\mathcal{L}_T)_{\text{($\nabla \phi $})^2} g^{\mu \nu } \partial _{\nu }\phi \right)}{g}= \nonumber \\
c( - \frac{1}{2}     \text{($\nabla \phi $})^2- R \;g(R)+3  \nabla _{\mu }\nabla ^{\mu }(\mathcal{L}_T)_R \nonumber \\
+ R \;(\mathcal{L}_T)_R-\frac{ \tau }{2}+2 V ), \label{eq:EoMCRGra}
\end{eqnarray}
where $c$ is a newly introduced proportional constant between the two equations of motion, Eqs. (\ref{eq:MGFE1}, \ref{eq:MGFE2}).
As mentioned in the previous section, the above equation is only valid under specific conditions (i.e., modeling constraints), which we need to infer next.

We assume a generic polynomial form of potential, 
\begin{equation}
	V=\sum_{n=1}^m \frac{\phi ^n \lambda _n(R)}{n!}, \label{eq:Vform}
\end{equation}
where $m$ denotes the highest power of $\phi$.

Refering to the previous section, the modeling constraints are derived by matching terms by terms of all orders of base field. 
By computing the term for the zeroth power of $\phi$ in Eq. (\ref{eq:EoMCRGra}), we obtain the following important constraint equation:
\begin{eqnarray}
 R \,g(R)+ \, \tau + c^{-1} \lambda _1(R)= \nonumber\\
 R^2 g'(R)+3 \nabla _{\mu }\nabla ^{\mu }\left(R\, g'(R)+g(R)\right). \label{eq:modelingGra}
\end{eqnarray}
One can solve for the gravitational model $\mathcal{L}_\text{gra} = \frac{R}{2} g(R)$, by solving $g(R)$ if $\tau$ and $\lambda _1$ are known.  We refer to this equation as the \textit{modeling equation for gravity}.  This is consistent with the general relativity that if we set $g(R)=1$ and $\lambda _1(R)=0$, the modeling equation yields the traced Einstein field equation $\tau + R = 0$.

By substituting the modeling equation for gravity into Eq. (\ref{eq:EoMCRGra}), for non-kinetic-$\phi$ terms, we obtain the following \textit{modeling equation for the scalar field}:
\begin{equation}
c \left(-6 \nabla _{\mu }\nabla ^{\mu }\partial _R\,V-2 R\, \partial _R\,V+4 V\right)+\lambda _1(R)=\partial _{\phi }V, \label{eq:modelingV}
\end{equation}
which (again) should only be solved for terms-by-terms of $\phi^r$, for an integer $r > 0$.

If we substitute the form of potential Eq. (\ref{eq:Vform}) into Eq. (\ref{eq:modelingV}), we obtain $\partial ^{\mu }  \lambda _n'(R)$.  This term is not trivial when solving Eq. (\ref{eq:modelingV}), because the constraint equations (for each term of $\phi^r$) are not regular differential equations for $\lambda_n(R)$.  However, if we impose the condition that, as part of the special configuration, the \textit{low-varying-$R$ condition} such that  $\partial ^{\mu }  \lambda _n'(R)$ is negligible, we obtain the equation 
\begin{equation}
\lambda _1(R)=\sum_{n=1}^m(\frac{2 c R\, \phi ^n \lambda _n'(R)}{n!}-\frac{4 c\, \phi ^n \lambda _n(R)}{n!}+\frac{n \phi ^{n-1} \lambda _n(R)}{n!}). \label{eq:lambdaEqn}
\end{equation}

The low-varying-$R$ condition leads to regular differential equations for the scalar model, this is, $\lambda_n(R)$; we refer to this model as the \textit{canonical model}.  Therefore, by definition, the canonical model requires a specific condition for the Ricci scalar; such a condition is the specific state implied by the mechanism.  We will justify this when we perform the conformal transformation to obtain model(s) in the Einstein frame in the next section(s), so that such specific condition for Ricci scalar is indeed only required in the Jordan frame.
The low-varying-$R$ condition also implies that we assume that all the kinetic terms of $\phi$ in Eq. (\ref{eq:EoMCRGra}) is neglectable.  When the field-map constraint Eq. (\ref{eq:fieldMapR}) applies, $\phi$ is one-to-one mapped onto $R$; therefore, the kinetic terms, such as $(\nabla \phi(R) )^2$, should be negligible because of the low-varying-$R$ condition.

 By re-ordering the dummy index in Eq. (\ref{eq:lambdaEqn}), we obtain the constraint equations for each term of $\phi^r$,
 \begin{equation}
\lambda _n'(R)=-\frac{\lambda _{n+1}(R)-4 c \lambda _n(R)}{2 c R}.
\end{equation}
Therefore, a coupled system of ordinary differential equations for $\lambda_n(R)$ is obtained.  In particular, we found a partial solution
\begin{align}
\lambda _m(R)=&R^2 c_m, \\
\lambda _{m-1}(R)=&R^2 c_{m-1}-\frac{R^2 c_m \log (R)}{2 c},
\end{align}
where $c_m$ is the integration constant(s).  If we use the recursive formula from the system of equations, we obtain two additional solutions of  $\lambda_n(R)$:
\small
\begin{align*}
\lambda _{m-2}(R) =& \\
  &R^2 (\frac{c_m \log ^2(R)}{8 c^2}  -\frac{c_{m-1} \log (R)}{2 c})+R^2 c_{m-2}, \\
\lambda _{m-3}(R) =&  \\
  R^2 (-\frac{c_m \log ^3(R)}{48 c^3}  &+\frac{c_{m-1} \log ^2(R)}{8 c^2}-\frac{c_{m-2} \log (R)}{2 c})+R^2 c_{m-3}. \label{eq:lambdaMminus}
\end{align*} \normalsize
One can express the general form of $\lambda_n$ as 
\begin{equation}
\lambda _n(R) =
R^2  P^{m-n}(\log (R)),
\end{equation}
where $P^{m-n}(z)$ denotes a polynomial with degree of $m-n$ of $z$.

After solving the canonical model for $\phi$, we attempt to solve the canonical model for gravity.  From the modeling equation for gravity, Eq. (\ref{eq:modelingGra}), we apply the low-varying-$R$ condition again, so the term
 $\nabla ^{\mu } (R  g'(R)+g(R))$ can be dropped.  We obtained a simple differential equation for the modeling of gravity:
\begin{equation}
c R \,g(R)+c\, \tau +\lambda _1(R)=c R^2 g'(R).
\end{equation}
If we assume that the matter term can be neglected, then the term $\tau$ is negligible, and we obtain the model of gravity:
\begin{equation}
\mathcal{L}_{\text{gra}}=\alpha _0+\frac{g_0 R^2}{2}+\frac{\alpha _1 R}{2},
\end{equation}
where we make use of the approximation of low-varying-$R$ condition so that we can approximate that $\lambda_1(R)$ is close to the linear order of $R$ so we expand $\lambda _1(R) \cong -c \left(4 \alpha _0+\alpha _1 R\right)$; $g_0$ is the integration constant, and $\alpha _0,\alpha _1$ are constants related to the expansion of $\lambda_1(R)$.
We show that the mechanism leads to the well-known $R^2$ gravity (or $R+R^2$ (Starobinsky) model\footnote{We denote $R^2$ gravity and $R+R^2$ theory the same model based on different conventions.}) plus the constant term. 
Interestingly, the Starobinsky model is a viable inflationary model in the Einstein frame after a conformal transformation \cite{1002.4928}.

In this section, we rely on the low-varying-$R$ condition to deduce the $R+R^2$ model.  Such a condition refers to a specific state of the modeling mechanism mentioned in Section 2, which appears too specific.  In the next section, we will show that models with the correspondence of general relativity (GR) can be generated; in the Einstein frame, the low-varying-$R$ condition is no longer required.  Therefore, the mechanism can lead to (re-produce) a general spacetime structure in the GR.

\begin{widetext}
Because we have solved the canonical model, the explicit form of a particular interesting case for the scalar field of the fourth power, that is, $m=4$, is found
\small
\begin{align}
 \frac{1}{2}   (\nabla \phi)^2
-\frac{1}{2} \phi ^2 R^2\left( \frac{c_4 \log ^2(R)}{8 c^2}-\frac{c_3 \log (R)}{2 c}+c_2 \right)- \nonumber\\
\phi R^2 \left( -\frac{c_4 \log ^3(R)}{48 c^3}+\frac{c_3 \log ^2(R)}{8 c^2}-\frac{c_2 \log (R)}{2 c}+c_1 \right)-\frac{1}{6} \phi ^3 R^2 \left(c_3 -\frac{c_4  \log (R)}{2 c}\right)-\frac{1}{24} c_4 R^2 \phi ^4+\frac{1}{2} R \left(\alpha _1+g_0 R+\frac{2 \alpha _0}{R}\right). \label{eq:m4Model}
\end{align}
\end{widetext}
\normalsize
Note the general statement of the mechanism is not limited to the model with $m=4$.

There are other viable  solutions (models) compared to the canonical model.  For example, if the following condition is satisfied,
\begin{equation}
\nabla _{\mu }\nabla ^{\mu }\left(R g'(R)+g(R)\right) = F(R, g(R), g'(R)),
\end{equation}
where $F$ is an analytical function, then the modeling equation for gravity Eq. (\ref{eq:modelingGra}) can be solved.  Such models are beyond the scope of the present study.

In the middle of this section, we assume the low-varying-$R$ condition to deduce the canonical model.  Such an assumption can be justified if there is a natural fixed point of $R$ during a specific state of the modeling mechanism.  We found it does exist.  
When the field-map constraint Eq. (\ref{eq:fieldMapR}) applies, $\phi$ is one-to-one mapped onto $R$; therefore, to obtain the low-varying-$R$ condition naturally, we look for the condition of low-varying-$\phi$ condition.  The trivial condition is that the scalar field $\phi$ stays at the minimum point of potential $V$.
By $\frac{d V}{d \phi} = 0$ from the potential at Eq. (\ref{eq:m4Model}) and substituting the field-map constraint Eq. (\ref{eq:fieldMapR}), we found the polynomial of $\log(R)$,
\begin{widetext}
\small
\begin{align}
8 c^3 \left(3 c_0 (c_4 v+c_3) \log ^2\left(R_0\right)-3 c_0^2 (v (c_4 v+2 c_3)+2 c_2) \log \left(R_0\right)+c_0^3 \left(v \left(c_4 v^2+3 c_3 v+6 c_2\right)+6 c_1\right)-c_4 \log ^3\left(R_0\right)\right)+ \nonumber \\
12 \left(2 c-c_0\right) c^2 \log (R) \left(-2 c_0 (c_4 v+c_3) \log \left(R_0\right)+c_0^2 (v (c_4 v+2 c_3)+2 c_2)+c_4 \log ^2\left(R_0\right)\right)+ \nonumber\\
6 c \left(c_0-2 c\right){}^2  \log ^2(R) \left(c_0 (c_4 v+c_3)-c_4 \log \left(R_0\right)\right)+\left(2 c-c_0\right){}^3 c_4 \log ^3(R)=0
\end{align}
\normalsize
If we further choose the choice of integration constants of $R_0$ and $v$ from Eq. (\ref{eq:fieldMapR}) such that $R=R_0$ when $\phi=v$, where $v$ is defined as the minimum point of $V$, we can further simplify the polynomial of $\log(R)$ above by substituting $R=R_0$ to yield
\small
\begin{align}
c_0 \left(8 c^3 \left(v \left(c_4 v^2+3 c_3 v+6 c_2\right)+6 c_1\right)-12 c^2 (v (c_4 v+2 c_3)+2 c_2) \log \left(R_0\right)+6 c (c_4 v+c_3) \log ^2\left(R_0\right)-c_4 \log ^3\left(R_0\right)\right)=0
 \label{eq:polylogR}. 
\end{align}
\normalsize
The specific value(s) of $R_0$ are the conditions of the modeling mechanism, that is, $R$ is close to the value(s) in the state of modeling.
\end{widetext}

\section{Generating Inflaton Model}

In the previous section, we derived the gravity and scalar models using the mechanism in the canonical case.  The mechanism introduced in Section 2 leads to the speculation of a specific state during the modeling mechanism.  To study this speculation, we applied the field-map constraint, Eq. (\ref{eq:fieldMapR}) to substitute the scalar field $\phi$ in the Lagrangian $\mathcal(R,\phi)$ to obtain a \textit{converging model}, denoted by $\tilde{\mathcal{L}}(R)$, of the mechanism, and then make use of conformal transformation.  We found that the mechanism generates an inflaton field in the Einstein frame (EF), that is, the inflaton model with GR correspondence.  (In this section, we study the case of $m=4$.  However, the general treatment is the same for cases where $m \ne 4$.)

Details of computational steps are presented in this section. We first apply the conditions of canonical models in the previous section: the matter sector is negligible, and the low-varying-$R$ condition is valid.  
Following the mechanism described in Section 2, the converging model $\tilde{\mathcal{L}}$ is obtained by substituting the field-map constraint $\phi(R)$ into the model in Eq. (\ref{eq:m4Model}),
so, the kinetic term, $(\nabla \phi(R) )^2$, should also be negligible because of the low-varying-$R$ condition.  Therefore, in such a specific state, $\tilde{\mathcal{L}}$ is effectively only a function of $R$ only, and the low-varying-$R$ condition applies.  
After the conformal transformation, we use a large-field approximation, which is justified by the work of Linde \cite{Linde_2018}, to yield an effective inflaton model.  Finally, we probe the parameters of such a model and compute the e-fold, which is consistent with observations.

Following the previous section, by assuming the low-varying-$R$ condition, the matter sector is negligible in the canonical model, and substituting the field-map constraint $\phi(R)$ from Eq. (\ref{eq:fieldMapR}), we obtained a simple form of the converging model from Eq. (\ref{eq:m4Model}),
\begin{align}
\tilde{\mathcal{L}}(R) = \alpha _0 +
\frac{\alpha _1 R}{2} + \frac{g_0 R^2}{2} + R^2 Q(\log (R)), \label{eq:LagEFInf}
\end{align}
where we introduce $Q(z)$ as a polynomial of $z$.  The additional terms to $R+R^2$ gravity are related to the terms from $Q$ polynomial.

By the formulas of the conformal transformation of $f(R)$ formalism \cite{1108.6266}, 
\begin{align}
	\varphi = \sqrt{\frac{3}{2}} \log (\frac{ f'(R) }{\sqrt{2 \pi }}), \\
	\mathcal{L}_{\text{EF}} = 
	\frac{\bar{R}}{2}-e^{-2 \sqrt{\frac{2}{3}} \varphi } (\frac{1}{2} e^{\sqrt{\frac{2}{3}} \varphi } R(\varphi )-f(\varphi ))-
	\frac{(\overset{\_ }{\nabla }\varphi )^2}{2}, \label{eq:LagEF}
\end{align}
where $R(\varphi)$ is obtained by the first equation above; $\tilde{\mathcal{L}}(R) =f(R)=f(\varphi)$; $\bar{R}$, $\overset{\_ }{\nabla }$ and $\mathcal{L}_{\text{EF}}$  denote the Ricci scalar, covariant derivative, and the model in Einstein frame respectively;
we recover general relativity (in Einstein frame), with a newly generated scalar field $\varphi$.
$R(\varphi)$ is approximately solved as 
\begin{equation}
	R(\varphi ) \simeq \frac{\eta+e^{\sqrt{\frac{2}{3}} \varphi }}{\kappa}, \label{eq:RMap}
\end{equation}
where $\eta, \kappa$ are constants, and we applied linear approximation, that is, $f'(R)$ expansion about $R - R_0$ up to first order\footnote{Such expansion is justified because of $R$ is around $R_0$ in the low-varying-R condition in Jordan frame.  $\eta, \kappa$ are constants constructed from polynomial $Q$.}; the model of Eq. (\ref{eq:LagEF}), and $R(\varphi)$ in the above form are general for cases with $m \ne 4$.

The model ($\mathcal{L}_\varphi$) of the scalar field  $\varphi$, obtained from Eq. (\ref{eq:LagEF}), becomes complicated after substituting $R(\varphi)$ and $f(\varphi)$.  We can apply the approximation to formulate an effective model that is valid for a large value of $\varphi$ owing to factors such as $e^{-\sqrt{\frac{2}{3}} \varphi }$ (as a large-field inflation model).
Large-field inflation models have been commonly studied.
The factor $e^{-\sqrt{\frac{2}{3}} \varphi }$, acts as a suppression factor for most terms of $\mathcal{L}_\varphi$.
\footnote{We also pick the leading order term for $\log(R)^{m}$.}
We obtain an effective model,
\begin{eqnarray*}
\mathcal{L}_{\text{EF}} = 
\frac{\bar{R}}{2}+\gamma _2 (\log (\eta +e^{\sqrt{\frac{2}{3}} \varphi }))^4-\gamma  \gamma _2-\frac{(\overset{\_}{\nabla} \varphi )^2}{2},
\end{eqnarray*}
where we introduce $\gamma$ and $\gamma_2$ to absorb lengthy constants.
We can probe the (orders of) parameters of the potential term of the scalar field $\varphi$ by observational constraints, namely, the tensor-to-scalar ratio $r$ and spectral index of scalar perturbations $n_s$.  
Note that the effective model is valid even for cases with $m \ne 4$, because Eq. (\ref{eq:LagEFInf}), Eq. (\ref{eq:LagEF}), and Eq. (\ref{eq:RMap}) are also valid for cases with $m \ne 4$. 

From \cite{Wang2014}, using the typical slow roll parameters, $\epsilon_V ,\eta_V$, to compute the tensor-to-scalar ratio $r$ and spectral index of scalar perturbations $n_s$,
\footnote{We probe the parameters by $n_s=0.965$ and $r<0.056$ for numerical solving.}
we probe the model's parameters of potential of scalar field $\varphi$ against the data from \cite{planck2018},
\begin{equation}
	\gamma \simeq \mathcal{O}(10^4), \eta \simeq \mathcal{O}(10^4),,
\end{equation}
and the range of field $\varphi$ during the inflation is
\begin{equation}
	\varphi \simeq 10.45 \to 15.99,  
\end{equation}
where we use the condition ($\epsilon_V=1$) to determine the upper range (i.e., $\varphi_\text{end}$).  This range justifies the assumption of a large-value approximation of the model.  We also found that the slow roll parameters at the crossing and e-folds are
\begin{equation}
	\epsilon_V\simeq0.0003, \eta_V \simeq -0.017, N_* \simeq 61,
\end{equation}
which are consistent with observations \cite{planck2018}.

In this section, 
we speculate that the model of the scalar field is \textit{absorbed} by the gravity model to yield the converging model, $\tilde{\mathcal{L}}$ (in the Jordan frame), because 
the additional terms in Eq. (\ref{eq:m4Model}) to $R+R^2$ model generated from the mechanism originates from the scalar field after substituting 
the field-map constraint - $\phi(R)$.
The converging model, $\tilde{\mathcal{L}}$, leads to a slow-roll scalar model of inflation with GR correspondence after the conformal transformation.  We speculate that the specific state of the modeling mechanism generates the converging model in the Jordan frame, which triggers inflation in the Einstein frame.
Finally, we show that the parameters of inflaton model can be probed and are consistent with observations \cite{planck2018}.   

In Ref. \cite{1108.6266}, it studies the (physical) equivalence of Jordan and Einstein frames.  Dicke argued that both frames are equivalent to the scaling of units of mass, length, and time.  We speculate that the Jordan frame might be more fundamental than the Einstein frame because the mechanism operates in the Jordan frame (JF) and generates modified gravity, while the effect of modified gravity goes away in the Einstein frame (EF), with a new degree of freedom as the
scalar field is obtained.

\section{Generating Dark Matter Model}

The remarkable success of the cold dark matter (CDM) model explains large-scale observations such as the one of the cosmic microwave background (CMB) and the abundance of light elements produced by big-bang nucleosynthesis (BBN). From Ref. \cite{1310.6061}, some of the cosmological predictions for CDM are in apparent conflict with some observations (e.g., cuspy-cored halos and an overabundance of satellite dwarf galaxies). A viable alternative model is scalar (field) dark matter (SFDM), which is described by the Klein-Gordon and Einstein field equations. The authors show that SFDM is compatible with observations of the evolving background universe to match observations of CMB and BBN (SFDM behaving as collision less on dust).  

In this section, we show that the theory from the mechanism can be interpreted as GR with a scalar field governed by the curved Klein-Gordon (KG) equation in the Einstein frame.  In the non-relativistic limit, KG equation in the GR is reduced to a system governed by the Schrodinger-Newton equation \cite{PhysRevD.69.124033}. 
On a scale larger than galaxies, dark matter behaves like cold dark matter;  at the scale below, the quantum mechanical nature suppresses dark matter structure formation owing to the minimum length scale \cite{0801.1442}. 
This property may naturally avoid cusp problem in the CDM model.
The observational constraints of the cosmic microwave background were consistent with the SFDM model \cite{1310.6061}.
Therefore, we speculate that the scalar dark matter model might fit both the observations of local measurements and large-scale measurements such as CMB. 

In the previous section, we generated the inflation model (in the Einstein frame) using the converging model (in the Jordan frame).  Here, 
we assume that the model generated from the mechanism (Eq. (\ref{eq:m4Model})) is valid after the inflation phase (phase corresponding to the converging model).
We study a trivial condition for the model of the scalar field $\phi$, Eq. (\ref{eq:m4Model}): defining $v$ such that as $\phi \simeq v$ corresponding to $R \simeq R_0$ around the low-varying-$R$ condition according to field-map constraint, Eq. (\ref{eq:fieldMapR}).
By substituting $\phi=v$, the scalar-field model is approximately,
\begin{equation}
	\mathcal{L}_\phi \simeq - V(R_0,v),
\end{equation}
where $V(R,\phi)$ is obtained from Eq. (\ref{eq:m4Model}).\footnote{The gravity sector is still the same $R+R^2$ model in Eq. (\ref{eq:m4Model}), with the condition that $R \simeq R_0$.}  
We yield a simplified model for the JF:
\begin{equation}
	\mathcal{L}_{\text{JF}} = \alpha _0+\alpha _2+\frac{g_0 R^2}{2}+\frac{\alpha _1 R}{2}, \label{eq:JFModelDM}
\end{equation}
where $\alpha _2 = -V(R_0, v)$.  Obviously, it is just $R + R^2$ model with a shifted-constant term.

We applied conformal transformation using the equations in Ref. \cite{1108.6266} yields the generated scalar field $\varphi$ in the Jordan frame as follows:
\begin{equation}
	\varphi =\sqrt{\frac{3}{2}} \log \frac{| R\, g_0+\frac{\alpha _1}{2}| }{\sqrt{2 \pi }}.
\end{equation}
We obtain the model in the Einstein frame,
\begin{equation}
	\mathcal{L}_{\text{EF}}  = -\frac{\bar{R}}{2}+a\, e^{-2 \sqrt{\frac{2}{3}} \varphi }+b\, e^{-\sqrt{\frac{2}{3}} \varphi }+g_1+\frac{(\overset{ \_ }{\nabla }\varphi )^2}{2},
\end{equation}
where $a=\alpha _0+\alpha _2-\frac{\alpha _1^2}{8 g_0}$, $b=-\frac{\alpha _1}{4 g_0}$, and $g_1=\frac{\sqrt{2 \pi }+2 \pi }{2 g_0}$.

If we define a new field as $e^{-\sqrt{\frac{2}{3}} \varphi }=\Phi$, the model in EF as
\begin{equation}
	\mathcal{L}_{\text{EF}}  = -\frac{\bar{R}}{2}+a \, \Phi ^2+b \, \Phi +g_1+\frac{3 (\overset{ \_ }{\nabla }\Phi )^2}{4 \Phi ^2}.
\end{equation}
We obtained the canonical form of the scalar field with a pre-factor term by further defining $\Phi =\rho +w$,
\begin{equation}
	\mathcal{L}_{\text{EF}}  = \frac{3}{2 (\rho +w)^2} \left(-\frac{\bar{R}_{\text{EF}}}{2}+\frac{(\overset{ \_ }{\nabla }\rho )^2}{2} + V_2(\rho )\right),
\end{equation}
where $\bar{R}=\frac{3 \bar{R}_{\text{EF}}}{2 (\rho +w)^2}$, $\bar{R}_{\text{EF}}$ denotes the effective Ricci scalar in the Einstein frame (to be justified next), and the potential is
\begin{equation}
	V_2(\rho )=\frac{2}{3} (\rho +w)^2 ((\rho +w) (a (\rho +w)+b)+g_1).
\end{equation}

We found the potential $V_2(\rho)$ can be casted into standard $\Phi^4$ model,
\begin{equation}
	V_2 = \frac{m_0^2 \Phi ^2}{2} + \frac{\sigma  \Phi ^4}{4},
\end{equation}
if
\begin{eqnarray}
	m_0=\frac{\sqrt{\frac{2}{3} \left(\sqrt{2 \pi }+2 \pi \right)}}{\sqrt{g_0}}, \sigma = -\frac{2 \left(\sqrt{2 \pi }+2 \pi \right)}{3 g_0 w^2}, \\
	\alpha _1=0,\alpha _0=\frac{-4 \alpha _2 g_0 w^2-\sqrt{2 \pi }-2 \pi }{4 g_0 w^2}, \label{eq:Phi4Conditions}
\end{eqnarray}
where $\frac{m_0}{\sqrt{-\sigma }} = w$.
The potential form above means $\Phi$ being staying to the minimum of $V_2$ if $\rho \ll w$.  Therefore, the pre-factor term of the potential is that, $ \frac{3}{2 (\rho +w)^2} \approx  \frac{3}{2 (w)^2}$; we obtain the model with canonical form in Einstein frame,
\begin{equation}
	\mathcal{L}_{\text{EF}} = \frac{-3}{2 w^2} \left(\frac{\bar{R}_{\text{EF}}}{2}-\frac{(\overset{ \_ }{\nabla }\Phi )^2}{2}-V_2(\Phi )\right)
\end{equation}
The pre-factor constant can be absorbed by redefining the volume element $d^4x$ in the action.  Therefore, we obtain a scalar field $\Phi$ that remains at the minimum of the potential and a typical GR correspondence.  The condition of $\Phi$ staying at a minimum can be realized if the kinetic energy of the scalar field $\Phi$ is low, which is consistent with the cold dark matter paradigm if we interpret $\Phi$ as a wave-dark-matter model.

We deduced the Klein-Gordon (KG) field in the Einstein frame from the mechanism under trivial conditions.  We postulate that this is a viable wave-dark-matter model.
In the Newtonian limit, the spherically symmetric Einstein-Klein-Gordon system is reduced to a system governed by the Schrodinger-Newton equations \cite{PhysRevD.69.124033}.  From Refs. \cite{1310.6061,0801.1442}, the scalar dark matter model might fit the observations for both local and large-scale measurements, such as CMB.

There are a number of studies on the wave nature of DM, such as ultralight bosons as CDM, a soliton core comprising a coherent standing wave at the center of halos, and self-interfering waves \cite{0801.1442}.  Most recently, in Ref. \cite{2304.09895}, the authors suggested that the wavelike dark matter model can reproduce all aspects of the system\footnote{The system is HS0810+2554 mentioned in Ref. \cite{2304.09895}.} 
whereas particle DM often fails by the analysis of the observation of gravitational lensing.

It is noteworthy that the conditions in Eq. (\ref{eq:Phi4Conditions}) to reproduce the $\Phi^4$ model as scalar dark matter requires $\alpha _1=0$.  Therefore, the model of gravity in the Jordan frame described by Eq. (\ref{eq:JFModelDM}) from the mechanism is indeed $R^2$ gravity without Einstein's term (i.e., the linear $R$ term).

Theoretically, the modification of Einstein gravity at high energies might be required from the quantum approach - the non-unitary and non-renormalizability of Einstein-Hilbert action (EHA) \cite{0809.1653}.  
$R^2$ gravity introduces one scalar degree of freedom with a one-dimensional parameter $M$ (the same dimensional parameter $m_0$ in Eq. (\ref{eq:Phi4Conditions})).  It provides the simplest correction to EHA (by inclusion of four-derivative terms), and it is renormalizable\footnote{
In Ref. \cite{0809.1653}, if we assume spacetime is in four dimension, there is non-trivial four-derivative extension of Einstein gravity that is free of ghosts and phenomenological viable - it is the $R^2$ gravity.}.

We consider this coincidence (obtaining $R^2$ gravity in the Jordan frame) to the connection of the modification of Einstein gravity at high energies, as a further study of the mechanism.

\section{Speculation of Energy-Scales Hierarchy}

This is a short section in which we explore the possibility of the energy-scale hierarchy implied in the Jordan frame by the solution of the mechanism.

At the end of Section 3, when we generate the modified gravity model from the mechanism described by Eq. (\ref{eq:m4Model}), we obtain the specific value(s) of $R$ because the condition of the specific state of the modeling mechanism, $R$ is close to the solution(s) of the polynomial of Eq. (\ref{eq:polylogR}), $R_0$.

This polynomial implies that there are $m-1$ possible solutions for $\log(R_0)$ in the Jordan frame.  
From Ref. \cite{Platania}, in the inflationary era, the energy scale $k$ was proposed to be connected to the Ricci scalar\footnote{In Ref. \cite{Platania}, the authors suggest the scale factor $a(t)$ is a power law that Ricci scalar is $R \sim t^{-2}$}
 as $k^2 \propto R$.
In addition, from Ref. \cite{1108.6266}, the Jordan frame is different from the Einstein frame by basic-unit rescaling.  Is a single energy scale in the Einstein frame related to multiple energy scales in the Jordan frame? 

Weinberg showed that inflation can be analyzed in the context of asymptotically safe theories of gravitation \cite{Weinberg_2010}.  This is natural, because in the early universe, asymptotical safety is important to the effect of short-distance physics.  Weinberg, Niedermaier, and Benedetti analyzed the asymptotic safety of the gravitational theory of $R + R^2$ and the term $R^{\mu\nu}\;R_{\mu\nu}$.  This theory shows that it is a possible renormalizable quantum theory of gravitation \cite{Weinberg_2010}.  It is clear $R^2$ gravity is theoretically interesting in the inflation era and is potentially renormalizable \cite{0809.1653}.

From Eq. (\ref{eq:polylogR}), we speculate that there are $m-1$ degenerated energy scales (described by the roots of the polynomial of $\log(R_0)^{m-1}$).
In the case of $m=4$, there may be a three-energy-scale hierarchy in the Jordan frame.  As we derive the $R^2$ gravity in the Jordan frame (by the mechanism), we consider studying the connection of short-distance physics/asymptotic safety to the energy scales hierarchy from the mechanism in future work.

\section{Conclusion}

In this work, motivated by observed low entropy, we introduce the low-entropy principle (section 2), and develop a modeling mechanism for modified gravity (section 3).  We found the specific model of gravity generated as the well-known $R^2$ gravity with additional logarithm terms.  By applying the conformal transformation, we generate both inflaton and dark matter in two different cases from the modified gravity in Einstein frame (section 4 and 5); therefore, the mechanism leads to the novel correspondence to general relativity in both inflation and dark matter paradigms. 

In section 6, We discuss the possibility that the mechanism implies the energy-scales hierarchy by the polynomial of $\log(R)$.  
It is worth to further studying such possibilities and potential connection to the observed energy-scales hierarchy from the standard model of particle physics. 

Furthermore, we found the specific condition of Ricci scalar, $R \simeq R_0$, from the modeling mechanism (section 3).  It is not a trivial spacetime condition.
Considering the work of asymptotic safety of the $R + R^2$ theory \cite{Weinberg_2010} and the generated $R^2$ gravity with additional logarithm terms (section 4), we speculate a connection of the modeling mechanism to a specific quantum gravity with a specific spacetime condition.

Finally, we compare Brans–Dicke theory versus our formalism.  
Brans–Dicke theory consists of an extra-degree of freedom \cite{brans2005}.  The main idea is to replace the gravitational constant by introducing a scalar field.  Our modeling mechanism does not assume a particular form of action; instead, we conjecture that nature picking the form(s) of action or models to respect the low-entropy principle.  Therefore, one can construct a general form of action from known formalisms of gravity theories and follow the modeling mechanism to obtain the \textit{preferred} model(s) of gravity.

\
\

\begin{acknowledgments}
JCHL would like to thank Professor Wang Yi of the Hong Kong University of Science and Technology for the valuable comments and advice.
JCHL performed the initial work of this study at Hong Kong University of Science and Technology.
\end{acknowledgments}

\bibliography{refs}

\begin{thebibliography}{24}%
\makeatletter
\providecommand \@ifxundefined [1]{%
 \@ifx{#1\undefined}
}%
\providecommand \@ifnum [1]{%
 \ifnum #1\expandafter \@firstoftwo
 \else \expandafter \@secondoftwo
 \fi
}%
\providecommand \@ifx [1]{%
 \ifx #1\expandafter \@firstoftwo
 \else \expandafter \@secondoftwo
 \fi
}%
\providecommand \natexlab [1]{#1}%
\providecommand \enquote  [1]{``#1''}%
\providecommand \bibnamefont  [1]{#1}%
\providecommand \bibfnamefont [1]{#1}%
\providecommand \citenamefont [1]{#1}%
\providecommand \href@noop [0]{\@secondoftwo}%
\providecommand \href [0]{\begingroup \@sanitize@url \@href}%
\providecommand \@href[1]{\@@startlink{#1}\@@href}%
\providecommand \@@href[1]{\endgroup#1\@@endlink}%
\providecommand \@sanitize@url [0]{\catcode `\\12\catcode `\$12\catcode
  `\&12\catcode `\#12\catcode `\^12\catcode `\_12\catcode `\%12\relax}%
\providecommand \@@startlink[1]{}%
\providecommand \@@endlink[0]{}%
\providecommand \url  [0]{\begingroup\@sanitize@url \@url }%
\providecommand \@url [1]{\endgroup\@href {#1}{\urlprefix }}%
\providecommand \urlprefix  [0]{URL }%
\providecommand \Eprint [0]{\href }%
\providecommand \doibase [0]{https://doi.org/}%
\providecommand \selectlanguage [0]{\@gobble}%
\providecommand \bibinfo  [0]{\@secondoftwo}%
\providecommand \bibfield  [0]{\@secondoftwo}%
\providecommand \translation [1]{[#1]}%
\providecommand \BibitemOpen [0]{}%
\providecommand \bibitemStop [0]{}%
\providecommand \bibitemNoStop [0]{.\EOS\space}%
\providecommand \EOS [0]{\spacefactor3000\relax}%
\providecommand \BibitemShut  [1]{\csname bibitem#1\endcsname}%
\let\auto@bib@innerbib\@empty
\bibitem [{\citenamefont {Smoot}(2007)}]{Smooth2007}%
  \BibitemOpen
  \bibfield  {author} {\bibinfo {author} {\bibfnamefont {G.~F.}\ \bibnamefont
  {Smoot}},\ }\bibfield  {title} {\bibinfo {title} {Nobel lecture: Cosmic
  microwave background radiation anisotropies: Their discovery and
  utilization},\ }\href {https://doi.org/10.1103/RevModPhys.79.1349} {\bibfield
   {journal} {\bibinfo  {journal} {Rev. Mod. Phys.}\ }\textbf {\bibinfo
  {volume} {79}},\ \bibinfo {pages} {1349} (\bibinfo {year}
  {2007})}\BibitemShut {NoStop}%
\bibitem [{\citenamefont {Wang}(2014)}]{Wang2014}%
  \BibitemOpen
  \bibfield  {author} {\bibinfo {author} {\bibfnamefont {Y.}~\bibnamefont
  {Wang}},\ }\bibfield  {title} {\bibinfo {title} {Inflation, cosmic
  perturbations and non-gaussianities},\ }\bibfield  {journal} {\bibinfo
  {journal} {Communications in Theoretical Physics}\ }\textbf {\bibinfo
  {volume} {62}},\ \href {https://doi.org/10.1088/0253-6102/62/1/19}
  {10.1088/0253-6102/62/1/19} (\bibinfo {year} {2014})\BibitemShut {NoStop}%
\bibitem [{\citenamefont {Akrami}\ \emph {et~al.}(2020)\citenamefont {Akrami}
  \emph {et~al.}}]{planck2018}%
  \BibitemOpen
  \bibfield  {author} {\bibinfo {author} {\bibfnamefont {Y.}~\bibnamefont
  {Akrami}} \emph {et~al.},\ }\bibfield  {title} {\bibinfo {title} {Planck 2018
  results},\ }\href {https://doi.org/10.1051/0004-6361/201833887} {\bibfield
  {journal} {\bibinfo  {journal} {A \& A}\ }\textbf {\bibinfo {volume} {641}}
  (\bibinfo {year} {2020})}\BibitemShut {NoStop}%
\bibitem [{\citenamefont {Li}\ \emph {et~al.}(2014)\citenamefont {Li},
  \citenamefont {Rindler-Daller},\ and\ \citenamefont {Shapiro}}]{1310.6061}%
  \BibitemOpen
  \bibfield  {author} {\bibinfo {author} {\bibfnamefont {B.}~\bibnamefont
  {Li}}, \bibinfo {author} {\bibfnamefont {T.}~\bibnamefont {Rindler-Daller}},\
  and\ \bibinfo {author} {\bibfnamefont {P.~R.}\ \bibnamefont {Shapiro}},\
  }\bibfield  {title} {\bibinfo {title} {Cosmological constraints on
  bose-einstein-condensed scalar field dark matter},\ }\bibfield  {journal}
  {\bibinfo  {journal} {Physical Review D}\ }\textbf {\bibinfo {volume} {89}},\
  \href {https://doi.org/10.1103/physrevd.89.083536}
  {10.1103/physrevd.89.083536} (\bibinfo {year} {2014})\BibitemShut {NoStop}%
\bibitem [{\citenamefont {Marsh}(2016)}]{1510.07633}%
  \BibitemOpen
  \bibfield  {author} {\bibinfo {author} {\bibfnamefont {D.~J.}\ \bibnamefont
  {Marsh}},\ }\bibfield  {title} {\bibinfo {title} {Axion cosmology},\ }\href
  {https://doi.org/10.1016/j.physrep.2016.06.005} {\bibfield  {journal}
  {\bibinfo  {journal} {Physics Reports}\ }\textbf {\bibinfo {volume} {643}},\
  \bibinfo {pages} {1} (\bibinfo {year} {2016})}\BibitemShut {NoStop}%
\bibitem [{\citenamefont {Broadhurst}\ \emph {et~al.}(2020)\citenamefont
  {Broadhurst}, \citenamefont {Martino}, \citenamefont {Luu}, \citenamefont
  {Smoot},\ and\ \citenamefont {Tye}}]{1902.10488}%
  \BibitemOpen
  \bibfield  {author} {\bibinfo {author} {\bibfnamefont {T.}~\bibnamefont
  {Broadhurst}}, \bibinfo {author} {\bibfnamefont {I.~D.}\ \bibnamefont
  {Martino}}, \bibinfo {author} {\bibfnamefont {H.~N.}\ \bibnamefont {Luu}},
  \bibinfo {author} {\bibfnamefont {G.~F.}\ \bibnamefont {Smoot}},\ and\
  \bibinfo {author} {\bibfnamefont {S.-H.~H.}\ \bibnamefont {Tye}},\ }\bibfield
   {title} {\bibinfo {title} {Ghostly galaxies as solitons of bose-einstein
  dark matter},\ }\bibfield  {journal} {\bibinfo  {journal} {Physical Review
  D}\ }\textbf {\bibinfo {volume} {101}},\ \href
  {https://doi.org/10.1103/physrevd.101.083012} {10.1103/physrevd.101.083012}
  (\bibinfo {year} {2020})\BibitemShut {NoStop}%
\bibitem [{\citenamefont {Hui}(2021)}]{2101.11735}%
  \BibitemOpen
  \bibfield  {author} {\bibinfo {author} {\bibfnamefont {L.}~\bibnamefont
  {Hui}},\ }\bibfield  {title} {\bibinfo {title} {Wave dark matter},\ }\href
  {https://doi.org/10.1146/annurev-astro-120920-010024} {\bibfield  {journal}
  {\bibinfo  {journal} {Annual Review of Astronomy and Astrophysics}\ }\textbf
  {\bibinfo {volume} {59}},\ \bibinfo {pages} {247} (\bibinfo {year}
  {2021})}\BibitemShut {NoStop}%
\bibitem [{\citenamefont {Lee}(2009)}]{0801.1442}%
  \BibitemOpen
  \bibfield  {author} {\bibinfo {author} {\bibfnamefont {J.-W.}\ \bibnamefont
  {Lee}},\ }\bibfield  {title} {\bibinfo {title} {Is dark matter a {BEC} or
  scalar field?},\ }\href {https://doi.org/10.3938/jkps.54.2622} {\bibfield
  {journal} {\bibinfo  {journal} {Journal of the Korean Physical Society}\
  }\textbf {\bibinfo {volume} {54}},\ \bibinfo {pages} {2622} (\bibinfo {year}
  {2009})}\BibitemShut {NoStop}%
\bibitem [{\citenamefont {Amruth}\ \emph {et~al.}(2023)\citenamefont {Amruth},
  \citenamefont {Broadhurst}, \citenamefont {Lim}, \citenamefont {Oguri},
  \citenamefont {Smoot}, \citenamefont {Diego}, \citenamefont {Leung},
  \citenamefont {Emami}, \citenamefont {Li}, \citenamefont {Chiueh},
  \citenamefont {Schive}, \citenamefont {Yeung},\ and\ \citenamefont
  {Li}}]{2304.09895}%
  \BibitemOpen
  \bibfield  {author} {\bibinfo {author} {\bibfnamefont {A.}~\bibnamefont
  {Amruth}}, \bibinfo {author} {\bibfnamefont {T.}~\bibnamefont {Broadhurst}},
  \bibinfo {author} {\bibfnamefont {J.}~\bibnamefont {Lim}}, \bibinfo {author}
  {\bibfnamefont {M.}~\bibnamefont {Oguri}}, \bibinfo {author} {\bibfnamefont
  {G.~F.}\ \bibnamefont {Smoot}}, \bibinfo {author} {\bibfnamefont {J.~M.}\
  \bibnamefont {Diego}}, \bibinfo {author} {\bibfnamefont {E.}~\bibnamefont
  {Leung}}, \bibinfo {author} {\bibfnamefont {R.}~\bibnamefont {Emami}},
  \bibinfo {author} {\bibfnamefont {J.}~\bibnamefont {Li}}, \bibinfo {author}
  {\bibfnamefont {T.}~\bibnamefont {Chiueh}}, \bibinfo {author} {\bibfnamefont
  {H.-Y.}\ \bibnamefont {Schive}}, \bibinfo {author} {\bibfnamefont {M.~C.~H.}\
  \bibnamefont {Yeung}},\ and\ \bibinfo {author} {\bibfnamefont {S.~K.}\
  \bibnamefont {Li}},\ }\bibfield  {title} {\bibinfo {title} {Einstein rings
  modulated by wavelike dark matter from anomalies in gravitationally lensed
  images},\ }\bibfield  {journal} {\bibinfo  {journal} {Nature Astronomy}\
  }\href {https://doi.org/10.1038/s41550-023-01943-9}
  {10.1038/s41550-023-01943-9} (\bibinfo {year} {2023})\BibitemShut {NoStop}%
\bibitem [{\citenamefont {Clifton}\ \emph {et~al.}(2012)\citenamefont
  {Clifton}, \citenamefont {Ferreira}, \citenamefont {Padilla},\ and\
  \citenamefont {Skordis}}]{1106.2476}%
  \BibitemOpen
  \bibfield  {author} {\bibinfo {author} {\bibfnamefont {T.}~\bibnamefont
  {Clifton}}, \bibinfo {author} {\bibfnamefont {P.~G.}\ \bibnamefont
  {Ferreira}}, \bibinfo {author} {\bibfnamefont {A.}~\bibnamefont {Padilla}},\
  and\ \bibinfo {author} {\bibfnamefont {C.}~\bibnamefont {Skordis}},\
  }\bibfield  {title} {\bibinfo {title} {Modified gravity and cosmology},\
  }\href {https://doi.org/10.1016/j.physrep.2012.01.001} {\bibfield  {journal}
  {\bibinfo  {journal} {Physics Reports}\ }\textbf {\bibinfo {volume} {513}},\
  \bibinfo {pages} {1} (\bibinfo {year} {2012})}\BibitemShut {NoStop}%
\bibitem [{\citenamefont {Stelle}(1977)}]{stelle1977}%
  \BibitemOpen
  \bibfield  {author} {\bibinfo {author} {\bibfnamefont {K.~S.}\ \bibnamefont
  {Stelle}},\ }\bibfield  {title} {\bibinfo {title} {Renormalization of
  higher-derivative quantum gravity},\ }\href
  {https://doi.org/10.1103/PhysRevD.16.953} {\bibfield  {journal} {\bibinfo
  {journal} {Phys. Rev. D}\ }\textbf {\bibinfo {volume} {16}},\ \bibinfo
  {pages} {953} (\bibinfo {year} {1977})}\BibitemShut {NoStop}%
\bibitem [{\citenamefont {Goldstein}\ \emph {et~al.}(2016)\citenamefont
  {Goldstein}, \citenamefont {Tumulka},\ and\ \citenamefont
  {Zangh\`{\i}}}]{PhysRevD.94.023520}%
  \BibitemOpen
  \bibfield  {author} {\bibinfo {author} {\bibfnamefont {S.}~\bibnamefont
  {Goldstein}}, \bibinfo {author} {\bibfnamefont {R.}~\bibnamefont {Tumulka}},\
  and\ \bibinfo {author} {\bibfnamefont {N.}~\bibnamefont {Zangh\`{\i}}},\
  }\bibfield  {title} {\bibinfo {title} {Is the hypothesis about a low entropy
  initial state of the universe necessary for explaining the arrow of time?},\
  }\href {https://doi.org/10.1103/PhysRevD.94.023520} {\bibfield  {journal}
  {\bibinfo  {journal} {Phys. Rev. D}\ }\textbf {\bibinfo {volume} {94}},\
  \bibinfo {pages} {023520} (\bibinfo {year} {2016})}\BibitemShut {NoStop}%
\bibitem [{\citenamefont {Yao}\ \emph {et~al.}(2021)\citenamefont {Yao},
  \citenamefont {Oliosi}, \citenamefont {Gao},\ and\ \citenamefont
  {Mukohyama}}]{PhysRevD.103.024032}%
  \BibitemOpen
  \bibfield  {author} {\bibinfo {author} {\bibfnamefont {Z.-B.}\ \bibnamefont
  {Yao}}, \bibinfo {author} {\bibfnamefont {M.}~\bibnamefont {Oliosi}},
  \bibinfo {author} {\bibfnamefont {X.}~\bibnamefont {Gao}},\ and\ \bibinfo
  {author} {\bibfnamefont {S.}~\bibnamefont {Mukohyama}},\ }\bibfield  {title}
  {\bibinfo {title} {Minimally modified gravity with an auxiliary constraint: A
  hamiltonian construction},\ }\href
  {https://doi.org/10.1103/PhysRevD.103.024032} {\bibfield  {journal} {\bibinfo
   {journal} {Phys. Rev. D}\ }\textbf {\bibinfo {volume} {103}},\ \bibinfo
  {pages} {024032} (\bibinfo {year} {2021})}\BibitemShut {NoStop}%
\bibitem [{\citenamefont {Harko}\ \emph {et~al.}(2013)\citenamefont {Harko},
  \citenamefont {Lobo},\ and\ \citenamefont {Minazzoli}}]{Harko_2013}%
  \BibitemOpen
  \bibfield  {author} {\bibinfo {author} {\bibfnamefont {T.}~\bibnamefont
  {Harko}}, \bibinfo {author} {\bibfnamefont {F.~S.~N.}\ \bibnamefont {Lobo}},\
  and\ \bibinfo {author} {\bibfnamefont {O.}~\bibnamefont {Minazzoli}},\
  }\bibfield  {journal} {\bibinfo  {journal} {Physical Review D}\ }\textbf
  {\bibinfo {volume} {87}},\ \href {https://doi.org/10.1103/physrevd.87.047501}
  {10.1103/physrevd.87.047501} (\bibinfo {year} {2013})\BibitemShut {NoStop}%
\bibitem [{\citenamefont {Carroll}\ and\ \citenamefont
  {Chen}(2004)}]{carroll2004spontaneous}%
  \BibitemOpen
  \bibfield  {author} {\bibinfo {author} {\bibfnamefont {S.~M.}\ \bibnamefont
  {Carroll}}\ and\ \bibinfo {author} {\bibfnamefont {J.}~\bibnamefont {Chen}},\
  }\href@noop {} {\bibinfo {title} {Spontaneous inflation and the origin of the
  arrow of time}} (\bibinfo {year} {2004}),\ \Eprint
  {https://arxiv.org/abs/hep-th/0410270} {arXiv:hep-th/0410270 [hep-th]}
  \BibitemShut {NoStop}%
\bibitem [{\citenamefont {CARROLL}\ and\ \citenamefont
  {CHEN}(2005)}]{CARROLL_2005}%
  \BibitemOpen
  \bibfield  {author} {\bibinfo {author} {\bibfnamefont {S.~M.}\ \bibnamefont
  {CARROLL}}\ and\ \bibinfo {author} {\bibfnamefont {J.}~\bibnamefont {CHEN}},\
  }\bibfield  {title} {\bibinfo {title} {Does inflation provide natural initial
  conditions for the universe?},\ }\href
  {https://doi.org/10.1142/s0218271805008054} {\bibfield  {journal} {\bibinfo
  {journal} {International Journal of Modern Physics D}\ }\textbf {\bibinfo
  {volume} {14}},\ \bibinfo {pages} {2335–2339} (\bibinfo {year}
  {2005})}\BibitemShut {NoStop}%
\bibitem [{\citenamefont {Felice}\ and\ \citenamefont
  {Tsujikawa}(2010)}]{1002.4928}%
  \BibitemOpen
  \bibfield  {author} {\bibinfo {author} {\bibfnamefont {A.~D.}\ \bibnamefont
  {Felice}}\ and\ \bibinfo {author} {\bibfnamefont {S.}~\bibnamefont
  {Tsujikawa}},\ }\bibfield  {title} {\bibinfo {title} {f(r) theories},\
  }\bibfield  {journal} {\bibinfo  {journal} {Living Reviews in Relativity}\
  }\textbf {\bibinfo {volume} {13}},\ \href
  {https://doi.org/10.12942/lrr-2010-3} {10.12942/lrr-2010-3} (\bibinfo {year}
  {2010})\BibitemShut {NoStop}%
\bibitem [{\citenamefont {Linde}(2018)}]{Linde_2018}%
  \BibitemOpen
  \bibfield  {author} {\bibinfo {author} {\bibfnamefont {A.}~\bibnamefont
  {Linde}},\ }\bibfield  {title} {\bibinfo {title} {On the problem of initial
  conditions for inflation},\ }\href
  {https://doi.org/10.1007/s10701-018-0177-9} {\bibfield  {journal} {\bibinfo
  {journal} {Foundations of Physics}\ }\textbf {\bibinfo {volume} {48}},\
  \bibinfo {pages} {1246–1260} (\bibinfo {year} {2018})}\BibitemShut
  {NoStop}%
\bibitem [{\citenamefont {Capozziello}\ and\ \citenamefont
  {Laurentis}(2011)}]{1108.6266}%
  \BibitemOpen
  \bibfield  {author} {\bibinfo {author} {\bibfnamefont {S.}~\bibnamefont
  {Capozziello}}\ and\ \bibinfo {author} {\bibfnamefont {M.~D.}\ \bibnamefont
  {Laurentis}},\ }\bibfield  {title} {\bibinfo {title} {Extended theories of
  gravity},\ }\href {https://doi.org/10.1016/j.physrep.2011.09.003} {\bibfield
  {journal} {\bibinfo  {journal} {Physics Reports}\ }\textbf {\bibinfo {volume}
  {509}},\ \bibinfo {pages} {167} (\bibinfo {year} {2011})}\BibitemShut
  {NoStop}%
\bibitem [{\citenamefont {Guzm\'an}\ and\ \citenamefont {Ure\~na
  L\'opez}(2004)}]{PhysRevD.69.124033}%
  \BibitemOpen
  \bibfield  {author} {\bibinfo {author} {\bibfnamefont {F.~S.}\ \bibnamefont
  {Guzm\'an}}\ and\ \bibinfo {author} {\bibfnamefont {L.~A.}\ \bibnamefont
  {Ure\~na L\'opez}},\ }\bibfield  {title} {\bibinfo {title} {Evolution of the
  schr\"odinger-newton system for a self-gravitating scalar field},\ }\href
  {https://doi.org/10.1103/PhysRevD.69.124033} {\bibfield  {journal} {\bibinfo
  {journal} {Phys. Rev. D}\ }\textbf {\bibinfo {volume} {69}},\ \bibinfo
  {pages} {124033} (\bibinfo {year} {2004})}\BibitemShut {NoStop}%
\bibitem [{\citenamefont {Cembranos}(2009)}]{0809.1653}%
  \BibitemOpen
  \bibfield  {author} {\bibinfo {author} {\bibfnamefont {J.~A.~R.}\
  \bibnamefont {Cembranos}},\ }\bibfield  {journal} {\bibinfo  {journal}
  {Physical Review Letters}\ }\textbf {\bibinfo {volume} {102}},\ \href
  {https://doi.org/10.1103/physrevlett.102.141301}
  {10.1103/physrevlett.102.141301} (\bibinfo {year} {2009})\BibitemShut
  {NoStop}%
\bibitem [{\citenamefont {Bonanno}\ and\ \citenamefont
  {Platania}(2016)}]{Platania}%
  \BibitemOpen
  \bibfield  {author} {\bibinfo {author} {\bibfnamefont {A.}~\bibnamefont
  {Bonanno}}\ and\ \bibinfo {author} {\bibfnamefont {A.}~\bibnamefont
  {Platania}},\ }\bibfield  {title} {\bibinfo {title} {{Asymptotically safe
  inflation from quadratic gravity}},\ }\href
  {https://doi.org/10.22323/1.263.0159} {\bibfield  {journal} {\bibinfo
  {journal} {PoS}\ }\textbf {\bibinfo {volume} {CORFU2015}},\ \bibinfo {pages}
  {159} (\bibinfo {year} {2016})}\BibitemShut {NoStop}%
\bibitem [{\citenamefont {Weinberg}(2010)}]{Weinberg_2010}%
  \BibitemOpen
  \bibfield  {author} {\bibinfo {author} {\bibfnamefont {S.}~\bibnamefont
  {Weinberg}},\ }\bibfield  {title} {\bibinfo {title} {Asymptotically safe
  inflation},\ }\bibfield  {journal} {\bibinfo  {journal} {Physical Review D}\
  }\textbf {\bibinfo {volume} {81}},\ \href
  {https://doi.org/10.1103/physrevd.81.083535} {10.1103/physrevd.81.083535}
  (\bibinfo {year} {2010})\BibitemShut {NoStop}%
\bibitem [{\citenamefont {Brans}(2005)}]{brans2005}%
  \BibitemOpen
  \bibfield  {author} {\bibinfo {author} {\bibfnamefont {C.~H.}\ \bibnamefont
  {Brans}},\ }\href@noop {} {\bibinfo {title} {The roots of scalar-tensor
  theory: an approximate history}} (\bibinfo {year} {2005}),\ \Eprint
  {https://arxiv.org/abs/gr-qc/0506063} {arXiv:gr-qc/0506063 [gr-qc]}
  \BibitemShut {NoStop}%
\end{thebibliography}%

\end{document}